\documentclass[sigconf]{acmart}

\AtBeginDocument{%
  \providecommand\BibTeX{{%
    \normalfont B\kern-0.5em{\scshape i\kern-0.25em b}\kern-0.8em\TeX}}}

\setcopyright{iw3c2w3}

\copyrightyear{2021}
\acmYear{2021}
\setcopyright{iw3c2w3}
\acmConference[WWW '21]{Proceedings of the Web Conference 2021}{April 19--23, 2021}{Ljubljana, Slovenia}
\acmBooktitle{Proceedings of the Web Conference 2021 (WWW '21), April 19--23, 2021, Ljubljana, Slovenia}
\acmPrice{}
\acmDOI{10.1145/3442381.3450000}
\acmISBN{978-1-4503-8312-7/21/04}

\usepackage[subtle,lists=tight]{savetrees}

\usepackage{verbatim}
\usepackage{multirow}
\usepackage{subfigure}

\usepackage{grffile}
\usepackage{algorithm, algorithmic}
\usepackage{amsmath, amsthm}
\usepackage{color}

\newcommand{\detector}{\mathcal{F}_{det}}
\newcommand{\extractor}{\mathcal{F}_{ext}}
\newcommand{\target}{\mathcal{F}_{tgt}}
\newcommand{\unwatermarklabel}{{non}}
\newcommand{\watermarklabel}{{wm}}
\newcommand{\ftwm}{g_\watermarklabel}
\newcommand{\fext}{e}
\newcommand{\dist}{d}
\newcommand{\key}{k}
\newcommand{\fekey}{\key_{FE}}
\newcommand{\mekey}{\key_{ME}}
\newcommand{\pspace}{\mathbb{W}}

\newcommand{\bits}{\texttt{BITS}}
\newcommand{\img}{\texttt{IMG}}

\newcommand{\uchida}{Uchida et al.}
\newcommand{\deepsign}{DeepSigns}
\newcommand{\ours}{RIGA}

\makeatletter
\def\blfootnote{\xdef\@thefnmark{}\@footnotetext}
\makeatother

\begin{document}

\title{RIGA: Covert and Robust White-Box Watermarking \\ of Deep Neural Networks}

\author{Tianhao Wang}
\affiliation{%
  \institution{Harvard University}
  \city{Cambridge}
  \state{MA}
  \country{United States}
}
\email{tianhaowang@fas.harvard.edu}

\author{Florian Kerschbaum}
\affiliation{%
  \institution{University of Waterloo}
  \city{Waterloo}
  \state{ON}
  \country{Canada}
}
\email{florian.kerschbaum@uwaterloo.ca}

\begin{abstract}
Watermarking of deep neural networks (DNN) can enable their tracing once released by a data owner to an online platform.
In this paper, we generalize white-box watermarking algorithms for DNNs, where the data owner needs white-box access to the model to extract the watermark. 
White-box watermarking algorithms have the advantage that they do not impact the accuracy of the watermarked model.
We propose \textbf{R}obust wh\textbf{I}te-box \textbf{GA}n watermarking (\ours), a novel white-box watermarking algorithm that uses adversarial training. 
Our extensive experiments demonstrate that the proposed watermarking algorithm not only does not impact accuracy, but also significantly improves the covertness and robustness over the current state-of-art. 
\end{abstract}

\maketitle

\section{Introduction}
\label{sec:introduction}

With data becoming an asset, data owners try to protect their intellectual property. One such method is protecting publicly accessible machine learning models derived from this data, which recently receives more and more attention as many Machine Learning as a Service (MLaaS) web applications, e.g., the Model Zoo by Caffe Developers, and Alexa Skills by Amazon, are appearing.
However, a machine learning model is an asset in itself, and can be easily copied and reused. Watermarking \cite{katzenbeisser15} may enable data owners to trace copied models. Watermarking embeds a secret message into the cover data, i.e.,~the machine learning model, and the message can only be retrieved with a secret key.

In the recent past, several watermarking algorithms for neural networks have been proposed \cite{adi18,chen19,merrer2017adversarial,Rouhani,szyller19,Uchida, zhang18, li2019piracy}.
These algorithms can be broadly classified into \emph{black-box watermarks} and  \emph{white-box watermarks} (see our related work discussion in Section \ref{sec:related}).
A black-box watermark can be extracted by only querying the model (black-box access).
A white-box watermark needs access to the model and its parameters in order to extract the watermark.
Recent studies \cite{shafieinejad19} show that black-box watermarks \cite{adi18,chen19,merrer2017adversarial,szyller19,zhang18} necessarily impact the model's accuracy, since they modify the training dataset and hence modify the learned function. This nature, however, can be unacceptable for some safety-critical applications
such as cancer diagnosis \cite{esteva2017dermatologist,khan2001classification}. A misclassified cancer patient with potential health consequences versus a model owner's intellectual property rights may be a hard to justify trade-off. 

On the contrary, white-box watermarks can work without accuracy loss. White-box watermarking was first introduced by Uchida et al.~\cite{Uchida} and later refined by Rouhani et al.~(DeepSigns) \cite{Rouhani}. They embed watermark messages into model weights by using some special regularizers during the training process. 

One line of work studies the \emph{watermark detection attack} \cite{TWang, shafieinejad19}. Specifically, \citet{TWang} show that Uchida et al.'s watermarking algorithm modifies weights distribution and can be easily detected. \citet{shafieinejad19} propose a supposedly strongest watermark detection attack called property inference attack.
Another line of work that attracts more interest is on \emph{watermark removal attacks} \cite{TWang, shafieinejad19, aiken2020neural, liu2020removing, chen2019refit}. In particular, most of the existing watermark removal attacks target at black-box watermarks \cite{shafieinejad19, chen2019refit, aiken2020neural, liu2020removing}. 
For white-box watermarks, \citet{TWang} show that an overwriting attack can easily remove \uchida's watermark. However, many removal attacks targeted at black-box watermarks can also be easily adapted to white-box settings, especially for those based on fine-tuning. 

In this paper, we generalize the research on white-box watermarking algorithms. First, we propose a formal scheme to reason about white-box watermarking algorithms (Section \ref{sec:background}) -- encompassing the existing algorithms of Uchida et al.~\cite{Uchida} and DeepSigns \cite{Rouhani}.
Then, we propose \ours, a novel, improved watermarking scheme that is both hard to detect and robust against all the existing watermark removal attacks. 
To improve the covertness, we encourage the weights distributions of watermarked models to be similar to the weights distributions of non-watermark models. 
Specifically, we propose a watermark hiding technique by setting up an adversarial learning network -- similar to a generative adversarial network (GAN) -- where the training of the watermarked model is the generator and a watermark detector is the discriminator. Using this automated approach, we show how to embed a covert watermark which has no accuracy loss. Furthermore, we make our white-box algorithm robust to model transformation attacks.
We replace the watermark extractor -- a linear function in previous works \cite{Rouhani,Uchida} -- with a deep neural network. This change largely increases the capacity to embed a watermark. Our watermark extractor is trained while embedding the watermark. The extractor maps weights to random messages except for the watermarked weights. 


Finally, we combine the two innovations into a novel white-box watermarking algorithm for deep neural networks named as \ours. We survey various watermark detection and removal attacks and apply them on \ours. We show that \ours~ {\em does not impact accuracy}, {\em is hard to detect} and {\em robust against model transformation attacks}.
We emphasize that a white-box watermark that does not impact accuracy cannot possibly protect against model stealing and distillation attacks \cite{hinton2015,Juuti19,Orekondy17}, since model stealing and distillation are black-box attacks and the black-box interface is unmodified by the white-box watermark.
However, white-box watermarks still have important applications when the model needs to be highly accurate, or model stealing attacks are not feasible due to rate limitation or available computational resources.




\section{Related Work}
\label{sec:related}


Watermarking techniques for neural networks can be classified into black-box and white-box algorithms. 
A black-box watermark can be extracted by only querying the model (black-box access).
A white-box watermark needs access to the model and its parameters in order to extract the watermark.
In this paper we present a white-box watermarking algorithm.
The first white-box algorithm was developed by Uchida et al.~\cite{Uchida}. 
Subsequently, Rouhani et al.~\cite{Rouhani} presented an improved version. 
We generalize both algorithms into a formal scheme for white-box watermarking algorithms and present their details in Section \ref{sec:background}.

The first attack on Uchida et al.'s algorithm was presented by Wang and Kerschbaum \cite{TWang}. 
They show that the presence of a watermark is easily detectable and that it can be easily removed by an overwriting attack.

The first black-box watermarking algorithms using backdoors were concurrently developed by Zhang et al.~\cite{zhang18} and Adi et al.~\cite{adi18}. 
Shafieinejad et al.~show that these backdoor-based watermarks can be easily removed using efficient model stealing and distillation attacks \cite{shafieinejad19}.
Attacks with stronger assumptions \cite{chen2019refit,yang2019, liu2020removing, aiken2020neural, chen2019refit} have later confirmed this result.

There are other types of black-box algorithms. Chen et al.~\cite{chen19} and Le Merrer et al.~\cite{merrer2017adversarial} use adversarial examples to generate watermarks. 
Szyller et al.~\cite{szyller19} modifies the classification output of the neural network in order to embed a black-box watermark. 
All black-box algorithms are obviously susceptible to sybil attacks \cite{hitaj19attacks}, unless access to multiple, differently watermarked models is prevented.


\section{Background}
\label{sec:background}

This section provides a formal definition of white-box watermarking of deep neural networks.
We provide a general scheme that encompasses at the least the existing white-box neural network watermarking algorithms \cite{Uchida,Rouhani}.

\subsection{Deep Neural Networks}
In this paper, we focus on deep neural networks (DNNs).
A DNN is a function $\mathcal{F}: X \rightarrow Y$, where $X$ is the input space, usually $\mathbb{R}^m$, and $Y$ is the collection of classes. There is a target joint distribution $P_{X, Y}$ on $(X, Y)$ which we want to learn about. 
A DNN $\mathcal{F}$ has function parameters $w$, which is a sequence of adjustable values to enable $\mathcal{F}$ fitting a wide range of mappings.
The values $w$ are commonly referred as \textit{model parameters} or \textit{model weights}, which we will use interchangeably.
For an instance $x \in X$, we represent the output of neural network $\mathcal{F}$ as $\mathcal{F}(x; w)$.
Let $\pspace$ be the parameter space of $w$, i.e.~$w \in \pspace$. $\pspace$ is usually high-dimensional real space $\mathbb{R}^n$ for DNNs, where $n$ is the total number of model parameters.
The goal of training a DNN $\mathcal{F}$ is to let $\mathcal{F}$ approximate the conditional distribution $P_{Y|X}$ by updating $w$.
The training of DNNs is the process of searching for the optimal $w$ in the parameter space to minimize a function $\mathcal{E}_o: \pspace \rightarrow \mathbb{R}$, which is typically a categorical cross-entropy derived from training data $(X_{train}, Y_{train})$. $\mathcal{E}_o$ is commonly referred to as \textit{loss function}. 
The accuracy of $\mathcal{F}$ after training depends on the quality of the loss function $\mathcal{E}_o$, while the quality of $\mathcal{E}_o$ in turn depends on the quality of the training data.
The search for a global minimum is typically performed using a stochastic gradient descent algorithm.

To formally define training, we assume there exist three algorithms:

\begin{itemize}
	\item $\mathcal{E}_o \leftarrow \textbf{DesignLoss}(X_{train}, Y_{train})$ is an algorithm that outputs loss function $\mathcal{E}_o$ according to the available training data.
	\item $w_{i+1} \leftarrow \textbf{TrainBatch}(\mathcal{E}_o, w_i)$ is an algorithm that applies \textbf{one iteration} of a gradient descent algorithm to minimize $\mathcal{E}_o$ with the starting weights $w_i$, and outputs the resulting weights $w_{i+1}$. 
	\item $\mathcal{F} \leftarrow \textbf{Train}(\mathcal{E}_o, w_0)$ is an algorithm that applies $\textbf{TrainBatch}(\mathcal{E}_o, w_i)$ iteratively for multiple steps where in the $i$-th iteration the input $w_i$ is the $w_i$ returned from the previous iteration. The iterations will continue until specific conditions are satisfied (e.g. model accuracy does not increase anymore). 
	The algorithm outputs the final model and its parameters $w$.
	For simplicity in the following text, when the initial weights $w_0$ are randomly initialized, we omit argument $w_0$ and simply write $\textbf{Train}(\mathcal{E}_o)$.
\end{itemize}

A well-trained DNN model $\mathcal{F}$ is expected to learn the joint distribution $P_{X, Y}$ well. Given DNN $\mathcal{F}$ and loss function $\mathcal{E}_o$, we say $\mathcal{F}$ is \emph{ $\varepsilon$-accurate} if
$
\Pr_{(x, y) \sim P_{X, Y}}[\mathcal{F}(x; w) \ne y] < \epsilon
$
where $w$ is the trained parameter returned by $\textbf{Train}(\mathcal{E}_o)$.

A regularization term \cite{buhlmann2011statistics}, or \textit{regularizer}, is commonly added to the loss function to prevent models from overfitting.
A regularizer is applied by training the parameters using $\textbf{Train}(\mathcal{E}_o+\lambda \mathcal{E}_R)$ where $\mathcal{E}_R$ is the regularization term and $\lambda$ is a coefficient to adjust its importance.

\subsection{White-box Watermarking for DNN models}
Digital watermarking is a technique used to embed a secret message, \textit{the watermark}, into cover data (e.g.~an image or video). It can be used to provide proof of ownership of cover data which is legally protected as intellectual property.
In white-box watermarking of DNN models the cover data are the model parameters $w$. 
DNNs have a high dimension of parameters, where many parameters have little significance in their primary classification task.
The over-parameterization property can be used to encode additional information beyond what is required for the primary task.

A white-box neural network watermarking scheme consists of a message space $\mathbb{M}$ and a key space $\mathbb{K}$. It also consists of two algorithms:

\begin{itemize}
	\item $m \leftarrow \textbf{Extract}(w, \key)$ is a deterministic polynomial-time algorithm that given model parameters $w$ and (secret) extraction key $\key$ outputs extracted watermark message $m$.
	\item $\mathcal{F}_{\watermarklabel}, \key \leftarrow \textbf{Embed}(\mathcal{E}_o, w_0, m)$ is an algorithm that given original loss function $\mathcal{E}_o$, a watermark message $m$ and initial model weights parameters $w_0$ outputs model $\mathcal{F}_{\watermarklabel}$ including its parameters $w_{\watermarklabel}$ and the (secret) extraction key $\key$. 
	In some watermarking algorithms \cite{Rouhani,Uchida} $\key$ can be chosen independently of $\mathcal{E}_0$, $w_0$ and $m$ using a key generation function \textbf{KeyGen}.
	For generality, we combine both algorithms into one. In the following text, when $w_0$ is randomly initialized, we omit argument $w_0$ and simply write $\textbf{Embed}(\mathcal{E}_o, m)$. 
\end{itemize}

The extraction of watermarks, i.e.~algorithm $\textbf{Extract}(w, \key)$ usually proceeds in two steps: (a) feature extraction, and (b) message extraction. 
The extraction key is also separated into two parts $\key=(\fekey, \mekey)$ for each of the steps in $\textbf{Extract}$.  
First, given feature extraction key $\fekey$, features $q$ are extracted from $w$ by feature extraction function $\ftwm$:
$$
q \leftarrow \ftwm(w, \fekey)
$$
For example, in the simplest case, the feature $q$ can be a subset of $w$, e.g. the weights of one layer of the model, and $\fekey$ is the index of the layer. This step is necessary given the complexity of a DNN's structure.

Second, given message extraction key $\mekey$, the message $m$ is extracted from the features $q$ by message extraction function $\fext$:
$$
m \leftarrow \fext(q, \mekey)
$$
We will refer to $\fext$ as \textit{extractor} in the remaining text. 

Embedding of the watermark, i.e.~algorithm $\textbf{Embed}(\mathcal{E}_o, m)$ is performed alongside the primary task of training a DNN. 
First, a random key $\key=(\fekey, \mekey)$ is randomly generated. 
Embedding a watermark message $m \in \mathbb{M}$ into target model $\target$ consists of regularizing $\target$ with a special regularization term $\mathcal{E}_{\watermarklabel}$. 
Let $\dist: \mathbb{M} \times \mathbb{M} \rightarrow \mathbb{R}$ be a distance function measures the discrepancy between two messages. 
For example, when $m$ is a binary string of length $n$, i.e.~$\mathbb{M} \subseteq \{0,1\}^n$, $\dist$ can simply be bit error rate.
Given a watermark $m$ to embed, the regularization term is then defined as:
$$
\mathcal{E}_{\watermarklabel} = \dist(\fext(\ftwm(w, \fekey), \mekey), m)
$$
The watermarked model $\target$ with model parameters $w_{\watermarklabel}$ is obtained by the training algorithm $\textbf{Train}(\mathcal{E}_o + \lambda \mathcal{E}_{\watermarklabel})$. 



\subsection{Requirements}
\label{sec:requirements}

There are a set of minimal requirements that a DNN watermarking algorithm should fulfill:

\textbf{Functionality-Preserving}: The embedding of the watermark should not impact the accuracy of the target model: 
$$
\Pr_{(x, y) \sim P_{X, Y}}[\mathcal{F}(x; w_{\watermarklabel}) = y] \approx \Pr_{(x, y) \sim P_{X, Y}}[\mathcal{F}(x; w) = y]
$$
where $w_{\watermarklabel}$ is returned by $\textbf{Train}(\mathcal{E}_o + \lambda \mathcal{E}_{\watermarklabel})$ and $w$ is returned by $\textbf{Train}(\mathcal{E}_o)$. 

\textbf{Robustness}: For any model transformation (independent of the key $\key$, e.g. fine-tuning) mapping $w_{\watermarklabel}$ to $w'$, such that model accuracy does not degrade, the extraction algorithm should still be able to extract watermark message $m'$ from $w'$ that is convincingly similar to the original watermark message $m$, i.e. if
$$
\Pr_{(x, y) \sim P_{X, Y}}[\mathcal{F}(x; w') = y] \approx \Pr_{(x, y) \sim P_{X, Y}}[\mathcal{F}(x; w_{\watermarklabel}) = y]
$$
where $w'$ is obtained from a model transformation mapping, then 
$$\textbf{Extract}(w', \key) \approx \textbf{Extract}(w_{\watermarklabel}, \key)$$

A further requirement we pose to a watermarking algorithm is that the watermark in the cover data is covert.
This is a useful property, because it may deter an adversary from the attempt to remove the watermark, but it is not strictly necessary.

\textbf{Covertness}: 
The adversary should not be able to distinguish a watermarked model from a non-watermarked one. Formally, we say a watermark is covert if no polynomial-time adversary algorithm $\mathcal{A}$ such that:
\[
\Pr  \left[
\begin{array}{l}
\mathcal{F}_0 \leftarrow \textbf{Train}(\mathcal{E}_o); \mathcal{F}_1, \key \leftarrow \textbf{Embed}(\mathcal{E}_o, m); \\
b \stackrel{\$}{\leftarrow} \{ 0, 1 \}; \mathcal{A}(\mathcal{F}_b) = b
\end{array}
\right]  \gg \frac{1}{2}
\]
where $\mathcal{A}$ is polynomial in the number of bits to represent $k$. 

In the literature \cite{adi18,zheng19} further properties of watermarking algorithms have been defined. We review them and show that they are met by our new watermarking scheme 
in Appendix \ref{apdx:requirements}.

\newcommand{\fematrix}{Z_{FE}}
\newcommand{\mematrix}{Z_{ME}}

\subsection{Example of Watermarking Scheme}
\label{sec:uchida}
We now formalize \uchida's watermarking algorithm \cite{Uchida} with our general scheme. Since \deepsign~ \cite{Rouhani} is similar to \uchida~in many aspects, we refer to Appendix \ref{sec:rouhani} for its formal scheme. In Uchida et al.'s algorithm, the message space $\mathbb{M} = \mathbb{R}_{[0, 1]}^t$. 
A typical watermark message $m \in \mathbb{M}$ is a $t$-bit binary string. 
Both feature extraction key space $\mathbb{K}_{FE}$ and message extraction key space $\mathbb{K}_{ME}$ are matrix spaces. 
The features $q$ to embed the watermark into are simply the weights of a layer of the DNN, i.e.~$\ftwm$ is the multiplication of a selection matrix $\fematrix$ with the vector $w$ of weights. Hence the feature extraction key $\fekey = \fematrix$. 
The message extractor $\fext$ does a linear transformation over the weights $w_l$ of one layer using message extraction key matrix $\mekey=\mematrix$, 
and then applies sigmoid function to the resulting vector to bound the range of values. 
The distance function $\dist$ is the binary cross-entropy between watermark message $m$ and extracted message $\fext(\ftwm(w, \fematrix), \mematrix)$.
Formally, Uchida et al.'s watermarking scheme is defined as follows:

\begin{itemize}
	\item $\ftwm:\pspace \times \mathbb{K}_{FE} \rightarrow \pspace_{l}$ where $\ftwm(w, \fematrix) = \fematrix w = w_l$. \\
		$\fematrix$ is a $|w_l| \times |w|$ selection matrix with a $1$ at position $(i, 1)$, $(i + 1, 2)$ and so forth, and $0$ otherwise, where $i$ is the start index of a layer.
		$\pspace_{l}$ is the parameter space of the weights of the selected layer, which is a subspace of $\pspace$.
	\item $\fext: \pspace_{l} \times \mathbb{K}_{ME} \rightarrow \mathbb{M}$ where $\fext(w_{l}, \mematrix) = \sigma(\mematrix w_l)$. \\
		$\mematrix$ is a $t \times |w_l|$ matrix whose values are randomly initialized. $\sigma$ denotes sigmoid function. 
	\item  $\dist: \mathbb{M} \times \mathbb{M} \rightarrow \mathbb{R}_{+}$ where $\dist(m, y) = m \log(y)+(1-m)\log(1-y)$ and 
	$y = \fext(\ftwm(w, \fematrix), \mematrix)$.
\end{itemize}

\newcommand{\muwm}{\mu_{wm}}
\newcommand{\munon}{\mu_{non}}
\newcommand{\losso}{\mathcal{E}_{o}}
\newcommand{\losswm}{\mathcal{E}_{wm}}
\newcommand{\lossdet}{\mathcal{E}_{det}}

\section{Covert and Robust Watermarking}
\label{sec:newwatermark}
In this section, we present the design of our new watermarking algorithm, \ours, and analyze the source of its covertness and robustness.
\subsection{Watermark Hiding}
\label{sec:defensedesign}
The white-box watermarking algorithms summarized in Section \ref{sec:background} are based on regularization
$\textbf{Train}(\losso + \lambda \mathcal{E}_{\watermarklabel})$. 
As demonstrated by \citet{TWang}, this extra regularization term detectably changes the distribution of weights in the target layer of the target models (referred as \emph{weights distribution} in the remaining text), which makes watermark detection feasible. In this section, we propose a technique that provably increase the covertness of watermarks. 

Denote $\munon$ and $\muwm$ as the probability distribution of the weights vector outputted by $\textbf{Train}(\mathcal{E}_o)$ and $\textbf{Train}(\mathcal{E}_o + \mathcal{E}_{wm})$, respectively. Both of $\munon$ and $\muwm$ are therefore defined on parameter space $\pspace$. Intuitively, to make embedded watermarks difficult to be detected, we need to make $\muwm$ to be as close to $\munon$ as possible. To hide the watermark, we impose the constraint that the weights of a watermarked model should be similar to the weights of a typical non-watermarked model. We define a \emph{security metric} $\lossdet$ to evaluate the covertness of the obtained watermarked weights, which serves as an additional regularizer in the new loss function for watermark embedding: 
\begin{equation}
\label{eq:detectionloss}
    \losso + \lambda_1 \losswm + \lambda_2 \lossdet
\end{equation}
where the hyperparameters $\lambda_1$ and $\lambda_2$ control the tradeoff between different loss terms. 

Later in this section, we introduce the relationship between the new loss function and the covertness goal theoretically. Prior to that, we first present the overall structure of our approach. We first collect a batch of non-watermarked models and extract a sample of non-watermarked weights $w_{non}$. The proposed scheme consists of two components: 1) target model $\target$ to be embedded with watermark, and 2) detector model $\detector$ which tries to distinguish $\target$'s weights from non-watermarked models' weights. 
This detector guides the training process of $\target$, ensuring that the learned weights $w$ are similar to $w_{non}$. 

To calculate $\lossdet$, we measure the Earth Mover Distance (EMD) between $\munon$ and $\muwm$: 
\begin{equation}
\label{eq:lossdet-original}
\begin{split}
\lossdet &= \detector(\muwm, \munon; \theta) \\
&= 
\inf_{\gamma \in \tau(\muwm, \munon)}(\int_{\pspace \times \pspace}\lVert w - w_{non} \rVert d \gamma(w, w_{non}))
\end{split}
\end{equation}
where $\theta$ is the parameters of $\detector$, $\tau(\muwm, \munon)$ is the collection of all joint probability measures on $\pspace \times \pspace$ with two marginal distributions of $\muwm$ and $\munon$ respectively. We use the Euclidean distance $\lVert w - w_{non} \rVert$ between model parameters to implement EMD. Developing more fine-grained distance metrics between models is an interesting future work. 

Unfortunately, the computation of Equation \ref{eq:lossdet-original} is intractable.  Instead, we use the Kantorovich-Rubinstein Duality \cite{arjovsky17}:
\begin{equation}
\label{eq:lossdet-KR}
\begin{split}
&\lossdet = \detector(\muwm, \munon; \theta) \\
&= 
\frac{1}{K} \sup_{\lVert \detector \rVert_L \le K} E_{w_{non} \sim \munon}[ \detector(w_{non}; \theta)] - E_{w \sim \muwm}[ \detector(w; \theta)] \nonumber
\end{split}
\end{equation}
where our detector $\detector$ is taken from the set of all functions $\pspace \rightarrow \mathbb{R}$ satisfying the $K$-Lipschitz condition. 

We implement $\detector$ as another neural network to approximate the EMD.
The idea of this scheme stems from the training of GAN models \cite{goodfellow2014generative}. 
However, our proposed approach differs from GANs in that $\target$ does not need the input noise to generate an output. Instead, the generated sample is taken directly from $\target$ itself, i.e.~its weights $w$ after each iteration of update. 
Similar to the principle of GAN, $\target$ and $\detector$ can be regarded as playing the following two-player minimax game with function $V(w,~\theta)$: 
\begin{equation}
\begin{split}
\min_w \max_{\theta} V(w, \theta) &= E[\log \detector(w_{\unwatermarklabel}; \theta)] + E[\log (1-\detector(w; \theta))] \nonumber
\end{split}
\end{equation}
Hence, in addition to training the primary task ($\mathcal{E}_o$) and embedding the watermark message ($\mathcal{E}_{\watermarklabel}$), $w$ in $\target$ is also updated according to the above equation, accompanied by the training of $\detector$:
\begin{align}
\label{eq:protectUpdatetheta}
\hat \theta & = \max_{\theta} (\log \detector(w_{\unwatermarklabel}; \theta) + \log (1-\detector(w; \theta))) \\
\label{eq:protectUpdatew}
\hat w & = \min_{w} (\mathcal{E}_o(w)+\lambda_1 \mathcal{E}_{\watermarklabel}(w)-\lambda_2
\underbrace{
\log \detector(w; \theta))
}_{\mathcal{E}_{det}}
\end{align}
During training, we alternately train $\target$ and $\detector$ with Equations \ref{eq:protectUpdatew} and \ref{eq:protectUpdatetheta}. 
We summarize our proposed method in Algorithm \ref{alg:protection}. The efficiency of this approach has been demonstrated \cite{arjovsky17}. 
We clamp the model parameters to ensure bounded first-order derivatives and therefore the Lipschitz condition \cite{arjovsky17}.
In order to improve the detection accuracy of $\detector$ and hence the covertness of final watermarked weights, a feature selection for the input of $\detector$ is necessary. In our implementation, we sort the weights as the neural networks are invariant in terms of different permutations of neurons in a layer \cite{ganju2018property}. 

We note that our watermark hiding technique is compatible with all existing white-box watermarking algorithms (\uchida~and \deepsign). In Section \ref{sec:eval-covertness}, we 
perform ablation study of the effectiveness of this watermarking hiding technique.

\begin{algorithm}
	\renewcommand{\algorithmicrequire}{\textbf{Input:}}
	\renewcommand{\algorithmicensure}{\textbf{Output:}}
	\caption{Watermark Hiding}
	\label{alg:protection}
	\begin{algorithmic}[1]
		\REQUIRE neural network $\mathcal{F}$ with loss function $\mathcal{E}_o$ and randomly initialized parameters $w$; detector neural network $\detector$ with parameters $\theta$; a white-box watermark algorithm featured by regularizer $\mathcal{E}_{\watermarklabel}$; hyperparameter $\lambda$; non-watermarked models with trained weights $w_{\unwatermarklabel}$; error tolerance $\epsilon$.
		\ENSURE watermarked model $\target$ with watermarked weights $w_{\watermarklabel}$ where covert message $m$ can be extracted.
		\WHILE{$|\detector(w; \theta) - \detector(w_{\unwatermarklabel}; \theta)| > \epsilon$ and $\target$ is not $\epsilon$-accurate}
		\STATE $\theta \leftarrow$ \textbf{TrainBatch} based on Equation~\ref{eq:protectUpdatetheta}
		\STATE $w \leftarrow$
		\textbf{TrainBatch} based on Equation~\ref{eq:protectUpdatew}
		\ENDWHILE
		\STATE \textbf{return} $\target$
	\end{algorithmic}
\end{algorithm}

\subsection{Watermark Embedding}
\label{sec:newwatermarkdesign}

\newcommand{\mrandom}{m_{r}}



Recall that a white-box watermarking algorithm \textbf{Embed} consists of three functions: $\ftwm$ outputs the features of DNN to be watermarked, $\fext$ extracts the watermark message from the features, and $\dist$ measures the discrepancy between an extracted message and the target message $m$.
 
In our proposed watermark embedding algorithm, we change the extractor $\fext$ to be another neural network, which we refer to as $\extractor$ with model parameters $\theta$. 
To extract a watermark message, we use the feature vector $q$ extracted by $\ftwm$ as the input to $\extractor$, and compare the output to the watermark message $m$ using $\dist$. 
Due to the strong fitting ability of neural networks, $\extractor$ can map $q$ to a wide range of data types of watermark messages $m$, e.g.~a binary string or even a 3-channel image. For different types of $m$, we choose the appropriate $\dist$ accordingly. 
Formally, our newly proposed white-box watermarking algorithm is defined as follows:

\begin{itemize}
	\item $\ftwm:\pspace\times \mathbb{K} \rightarrow \pspace_{l}$ where $\ftwm(w, Z) = Zw$.\\ 
	    As in Uchida et al.'s scheme the $\ftwm$ outputs a layer of $w$.
	\item $\fext:\pspace_{l}\times \Theta$ where $\fext(w_l, \theta) = \extractor(w_l; \theta)$. \\
	    $\extractor$ is a DNN with parameter $\theta$, and $\Theta$ is $\extractor$'s parameter space. 
	\item $\dist: \mathbb{M} \times \mathbb{M} \rightarrow \mathbb{R}$ varies for different data types of $m$.\\
	For binary strings we use cross-entropy as in Uchida et al.'s scheme, for images we use mean squared error for pixel values.
\end{itemize}

Our new algorithm largely increases the capacity of the channel in which the watermark is embedded, and hence allows to embed different data types, including pixel images, whilst in both Uchida et al.'s algorithm and DeepSigns, the embedded watermarks are restricted to binary strings. In the embedding algorithms of those two previous schemes, the number of embedded bits should be smaller than the number of parameters $w$, since otherwise the embedding will be overdetermined and cause large embedding loss. However, in our new scheme, the adjustable parameter is not only $w$ but also $\theta$ of $\extractor$, which largely increases the capacity of the embedded watermark. 
  
Next to enabling larger watermarks, the increased channel capacity of a neural network extractor also enhances the robustness of the watermark. In Uchida et al.'s scheme, $\fext$ is the sigmoid of a linear mapping defined by a random matrix $X$, which can be viewed as a single layer perceptron. The resulting watermark can be easily removed by overwriting. As shown by Wang and Kerschbaum \cite{TWang}, to remove the original watermark binary string $m$ without knowing the secret (key) matrix $X$, an adversary can randomly generate a key matrix $X^{*}$ of the same dimension, and embed his/her own watermark $\mrandom$ into the target neural network. 
The result of $Xw$ is a simple linear projection of $w$. Because of the low capacity of a linear mapping, a randomly generated $X^{*}$ is likely to be very similar to the original embedding matrix $X$ in some of the \textit{watermark bit positions}. 
Hence an adversary who randomly embeds a new watermark is likely to overwrite and remove the existing watermark at those bit positions.

In our algorithm, we make the watermark extraction more complex and the secret key more difficult to guess by adding more layers to the neural network, i.e.~we replace the secret matrix $X$ in Uchida et al.'s scheme by a multi-layer neural network, $\extractor$.

In existing white-box watermarking algorithms, the extractor $\fext$ is a static, pre-determined function depending only on the key $\key$. In our scheme, however, $\extractor$ is trained alongside the target model to enable fast convergence. 
Instead of only training $\target$ and updating $w$, we update $w$ and $\theta$ alternately to embed watermark message $m$ into $\target$ in an efficient way, as summarized in the following equation:
\begin{equation}
 \label{eq:alternativelytrain}
 \hat w, \hat \theta = \min_{w, \theta}(\mathcal{E}_o(w) + \lambda \dist(m, \extractor(w, \theta)))
\end{equation}
The above equation implies that the extractor $\fext=\extractor(\cdot; \theta)$ adapts to the message $m$. If that is the only goal of $\extractor$, it will result in a trivial function that ignores the input and maps everything to the watermark message $m$. 
To ensure the extractor being a reasonable function, we collect non-watermarked weights $w_{\unwatermarklabel}$ and train $\extractor$ to map $w_{\unwatermarklabel}$ to some random message $\mrandom$ other than watermark message $m$. Hence our new parameter update equations are:

\begin{align}
	\label{eq:updateEmbedder}
	\begin{split}
	\hat \theta & = \min_{\theta} (\dist(m, \extractor(w; \theta)) + \dist(\mrandom, \extractor(w_{\unwatermarklabel}; \theta)))
	\end{split}\\
	\label{eq:updateTarget}
	\hat w & = \min_{w} (\mathcal{E}_o(w) + \lambda
	\underbrace{
	\dist(m, \extractor(w; \theta)))
	}_{\mathcal{E}_{\watermarklabel}}
\end{align}

Because of the adaptive nature of $\extractor$, one may worry that two model owners will obtain the same extractor $\fext$ if the watermark message they choose is the same. However, this is nearly impossible. $\extractor$ is not only adaptive to $m$, but also many other factors such as the watermarked features $q$. Even if all of the settings are the same, since the loss function of neural networks is non-convex and there are numerous local minima, it is almost impossible for two training processes to fall into the same local minimum. We experimentally validate our hypothesis in Appendix \ref{sec:integrity}.

\begin{algorithm}
	\renewcommand{\algorithmicrequire}{\textbf{Input:}}
	\renewcommand{\algorithmicensure}{\textbf{Output:}}
	\caption{Watermark Embedding}
	\label{alg:embedding}
	\begin{algorithmic}[1]
		\REQUIRE neural network $\mathcal{F}$ with loss function $\mathcal{E}_o$ and parameters $w$; extractor $\extractor$ with randomly initialized parameters $\theta$; watermark message $m$; hyperparameter $\lambda$; set of pre-trained non-watermarked models with weights $\{ w_{\unwatermarklabel} \}$; error tolerance $\epsilon$.  
		\ENSURE watermarked model $\target$; trained extractor $\extractor$.
        \STATE Generate random messages $\mrandom \stackrel{\$}{\leftarrow} \mathbb{M}$ for every $w_{\unwatermarklabel}$. 
		\WHILE{$\dist(\extractor(w; \theta), m) > \epsilon$ and $\target$ is not $\epsilon$-accurate}
			\STATE $\theta \leftarrow$ \textbf{TrainBatch} based on Equation~\ref{eq:updateEmbedder}
			\STATE $w \leftarrow$ \textbf{TrainBatch} based on Equation~\ref{eq:updateTarget}
		\ENDWHILE
		\STATE \textbf{return} $\target$, $\extractor$
	\end{algorithmic}
\end{algorithm}

\subsection{Combination}
\label{sec:comb}
The watermark hiding and embedding algorithms mentioned in the two previous sections are similar in the sense that they both use a separate deep learning model ($\detector$ and $\extractor$) to hide or embed watermarks into the target model. 
It is hence natural to combine the two algorithms into one new neural network watermarking scheme, which we name it \ours. 
The workflow of \ours~ is summarized in Appendix \ref{apdx:flowchart}. 
In each round of training, $\target$'s weights $w$ are updated by loss function
\begin{equation}
\label{eq:tgt_update}
\mathcal{E}_o + \lambda_1 \mathcal{E}_{\watermarklabel} + \lambda_2 \mathcal{E}_{det}
\end{equation}
and $\extractor$ and $\detector$ are updated using Equations \ref{eq:updateEmbedder} and \ref{eq:protectUpdatetheta}, respectively. 
Our newly proposed white-box watermarking scheme for deep neural networks has several advantages including that it \textit{does not impact model accuracy}, \textit{is covert} and \textit{robust against model modification attacks, such as overwriting}, as demonstrated in the evaluation section.

\section{Evaluation Setup}
\label{sec:experimentsettings}

\subsection{Evaluation Protocol}
We evaluate \ours's impact on model performance, as well as the covertness and robustness of \ours.
\footnote{RIGA's implementation is available at \url{https://github.com/TIANHAO-WANG/riga}}
We show that our watermark hiding technique significantly improves the covertness of white-box watermarks by evaluating against the watermark detection approaches proposed by \citet{TWang} and \citet{shafieinejad19}. We demonstrate the robustness of \ours~ by evaluating against three watermark removal techniques: overwriting \cite{TWang}, fine-tuning \cite{adi18, chen2019refit}, and weights pruning \cite{Uchida, Rouhani, liu2018fine}. For all attacks we evaluate, we consider the strongest possible threat models. 
We refer to Appendix \ref{sec:integrity} for results on evaluating \ours's validity. 

\subsection{Benchmark and Watermark Setting}
Table \ref{table:benchmark} summarizes the three benchmarks we use. We leave the description of datasets and models to the Appendix \ref{apdx:dataset} and \ref{apdx:model}, respectively. 
To show the general applicability of \ours, we embed watermarks into different types of network layers. 
For Benchmark 1, we embed the watermarks into the weights of the second to last fully-connected layer. 
For Benchmark 2, we embed the watermarks into the third convolutional layer of the Inception-V3. 
Because the order of filters is arbitrary for a convolutional layer, we embed the watermark message into the mean weights of a filter at each filter position. 
For Benchmark 3, the watermarks are embedded into the weights of~``input gate layer'' of the LSTM layer. 
We use simple 3-layer fully-connected neural networks as the architecture for both $\extractor$ and $\detector$ in our experiments.
We use two different data types of watermarks, including a \emph{256-bit random binary string} and a \emph{$128\times 128$ 3-channel logo image}. In the following text, we denote models watermarked with 256-bit binary string as \bits, and models watermarked with logo image as \img. For example, if a Benchmark 2's model watermarked with binary string, we denote it as Benchmark 2-\bits. 

We use Adam optimizer with the learning rate $10^{-4}$, batch size 100, $\beta_1 = 0.5$, and $\beta_2 = 0.999$ to train these networks. We train 500 non-watermarked models for each of the benchmark and extracts the sample of non-watermarked weights $w_{non}$ for implementing Algorithm \ref{alg:protection} and \ref{alg:embedding}. We set $\lambda_1 = 0.01$ and $\lambda_2=0.1$ in Equation \ref{eq:tgt_update}. 
For the GAN-like training, we set weights clipping limit to 0.01 to ensure the Lipschitz condition. 

\subsection{Watermark Evaluation Metrics}
\paragraph{Bit Error Rate (BER)} For binary string watermarks, we use bit error rate as the metric to measure the distance between original and extracted watermarks. BER is calculated as 
$
    \frac{1}{n} \sum_{i=1}^n I[|\hat m_i - m_i| > 0.5]
$
where $\hat m$ is the extracted watermark, $I$ is the indicator function, and $n$ is the size of the watermark. 

\paragraph{Embedding Loss} BER may not reflect the exact distance between original and extracted watermarks, as both $\hat m_i=0.99$ and $\hat m_i=0.51$ will be interpreted as ``correct bit" for $m_i = 1$. Therefore we introduce Embedding Loss as an alternate metric for binary string watermarks, which computes the \emph{binary cross entropy} between $\hat m$ and $m$. For image watermarks, we use \emph{L2 distance} between original and extracted image as the embedding loss. 

\subsection{Watermark Detection Attack}
\citet{TWang} report that the existing white-box watermarking algorithms change target model's weights distribution, which allows one to detect watermark simply by visual inspection. 
\citet{shafieinejad19} propose a more rigorous and supposedly strongest watermark detection attack called \emph{property inference attack}. The intuition of property inference attack is that the common patterns of watermarked weights can be easily learned by a machine learning model if enough samples of such weights distributions are provided.
Specifically, a property inference attack trains a watermark detector $\detector$ on weights distributions extracted from various neural network models that are either non-watermarked or watermarked. 
Our evaluation assumes the worst case where the attacker knows the training data, the exact model architecture of $\target$, and the feature extraction key $\fekey$. Note that this threat model is overly strong, but we consider the worst case here to demonstrate the effectiveness of our watermark hiding algorithm. 
We leave the details of property inference attack implementation to Appendix \ref{apdx:pi_attack-setup}. 

\subsection{Watermark Removal Attack}
\subsubsection{Overwriting Attack}
The attacker may attempt to remove the original watermark by embedding his/her own watermarks into the target model. This technique has been shown to be effective against \uchida's watermark \cite{TWang}. Specifically, for every epoch, the attacker generates a new watermark message $m$ and a secret key $\mekey$, embeds this message into the target model and hopes it overwrites the original watermark. In our experiments, we assume the worst case such that the attacker knows everything but the message extraction key $\mekey$. That is, the attacker has access to the original training set and has knowledge of the feature extraction key $\fekey$, which means that the attacker is aware of the layer in $\target$ where watermark message $m$ is embedded. 
The only thing the attacker does not know is the model parameters of $\extractor$, $\theta$, which serve as our message extraction key $\mekey$. 



\subsubsection{Fine-tuning Attack}
Fine-tuning is the most common watermark removal attacks considered in the literature \cite{Uchida, adi18, Rouhani, chen2019refit, liu2018fine}. Formally, for a trained model $\target$ with parameter $w$, we fine-tune the model by updating $w$ to be $\textbf{Train}(\mathcal{E}_{ft})$ where $\mathcal{E}_{ft}$ can be the same as or different from $\mathcal{E}_o$. 
Note that we train $\target$ without the watermarking-related regularizers ($\mathcal{E}_{\watermarklabel},~ \mathcal{E}_{det}$) during fine-tuning. \citet{adi18} propose several variants of fine-tuning process. In our evaluation, we always fine-tune the entire model because fine-tuning the output layers only is insufficient to remove white-box watermarks by design, as long as the watermarks are not embeded on the output layers. We evaluate both FTAL and RTAL processes in \citet{adi18}. Specifically, FTAL directly fine-tunes the entire model; when using RTAL, the output layer is randomly initialized before fine-tuning. 

The fine-tuning learning rates set in \citet{adi18} are the same as those in the original training process. \citet{chen2019refit} argue that the learning rates set in the fine-tuning process in \citet{adi18} are too small and propose a generic black-box watermark removal attack called REFIT by setting larger fine-tuning learning rates. Specifically, they set the initial fine-tuning learning rate to be much larger, e.g. 0.05. During the fine-tuning process, the learning rate is decayed by 0.9 for every 500 iteration steps. We evaluate REFIT on \ours~ with many different initial learning rates.   


\subsubsection{Weights Pruning Attack}
Weights pruning, i.e., removing connections between some neurons in the neural network, is another common post-processing operation of DNNs and hence a plausible threat to embedded watermarks. It has been verified that pruning is not effective on removing 
either \uchida~or \deepsign. For the completeness of the evaluation, we also test the robustness of \ours~against pruning. 

Previous work found that combining pruning-based techniques with fine-tuning could improve the effectiveness of black-box watermark removal \cite{liu2018fine}. However, as reported in \citet{chen2019refit}, pruning is not necessary with a properly designed learning rate schedule for fine-tuning even for black-box watermarks. Nevertheless, we also evaluate the performance of fine-pruning on \ours~ and present results in Appendix \ref{sec:fine-pruning}. 

\section{Evaluation Results}
\label{sec:eval-results}

\subsection{Model Performance}
\label{sec:eval-performance}
We expect the accuracy of the watermarked deep learning model not to degrade compared to the non-watermarked models. Table \ref{table:benchmark_acc} summarizes the mean and 95\% upper/lower bound of the accuracy of non-watermarked models and models with a watermark message embedded by \ours. The 95\% upper/lower accuracy bound is obtained by training over 1000 models with different random seeds. 
The results demonstrate that \ours~maintains model accuracy by optimizing the original objective function whilst simultaneously embedding a watermark message. 
In some cases, e.g. Benchmark 2, we even observe a slight performance improvement. This is because our two extra loss terms ($\mathcal{E}_{\watermarklabel}$ and $\mathcal{E}_{det}$) serve as regularizers whereas the non-watermarked models are trained without regularizers. Regularization, in turn, helps the model avoid overfitting by introducing a small amount of noise into the target model. Table \ref{table:benchmark_acc} also includes the values of the embedding loss for each benchmark after training, which are close to zero in all cases. 
We conclude that neural networks are capable of maintaining accuracy while memorizing information injected by regularization.

\begin{table*}[t]
\centering
\begin{tabular}{|c|c|c|c|c|}
\hline
\textbf{Benchmark ID} & \textbf{Dataset} & \textbf{Task} & \textbf{Watermark Embed Layer} & \textbf{Target Model}  \\ \hline
1 & MNIST      & Digit Recognition        & Fully-Connected               & LeNet                  \\ \hline
2 & CelebA     & Gender Classification       & Conv. Layer                    & Inception-V3           \\ \hline
3 & Amazon Fine Food  & Sentiment Classification    & LSTM                           & Embedding Layer + LSTM \\ \hline
\end{tabular}
\caption{Benchmark Setup}
\label{table:benchmark}
\end{table*}

\begin{table*}[t]
\centering
\begin{tabular}{|c|c|c|c|c|c|}
\hline
\multirow{2}{*}{\textbf{BM}} & \multirow{2}{*}{\textbf{Non-watermarked  Accuracy}} & \multicolumn{2}{c|}{\textbf{Watermarked   Model Accuracy}} & \multicolumn{2}{c|}{\textbf{Watermark Embedding Loss}} \\ \cline{3-6} 
                                  &                                               & \textbf{256-BITS}            & \textbf{Image}              & \textbf{256-BITS}                & \textbf{Image}                \\ \hline
1                             & 97.92\% (97.21\%, 98.32\%)                    & 97.8\% (96.97\%, 98.57\%)    & 97.32\% (97.22\%, 97.64\%)  & 2.62E-06                                & 5.12E-05                      \\ \hline
2                            & 95.05\% (94.74\%, 95.96\%)                    & 96.15\% (95.02\%, 96.72\%)   & 96.02\% (94.79\%, 96.63\%)  & 3.76E-06                                & 2.28E-04                      \\ \hline
3                & 83.89\% (82.72\%, 84.55\%)                    & 83.33\% (82.4\%, 84.32\%)    & 83.12\% (82.29\%, 84.01\%)  & 1.11E-06                                & 6.31E-06                      \\ \hline
\end{tabular}
\caption{Benchmark Accuracy Intervals and Watermark Embedding Loss}
\label{table:benchmark_acc}
\end{table*}

\subsection{Covertness}
\label{sec:eval-covertness}
\begin{figure*}
	\centering 
	\includegraphics[width=\textwidth]{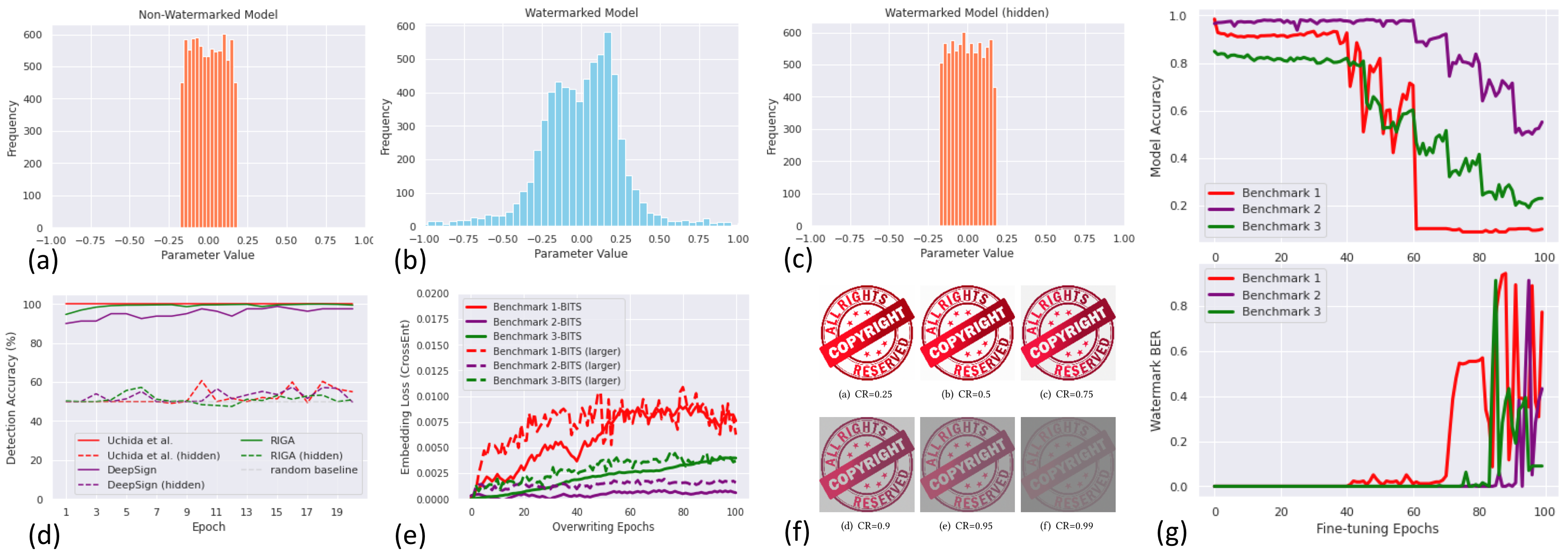}
	\caption{
	Results for (a)-(c) visualization of weights distribution, (d) property inference attacks, 
	(e) overwriting attacks, (f) weights pruning attacks on \img, where CR refers to compression ratio, (g) fine-tuning attacks.
	}
	\label{fig:huge} 
\end{figure*}

Figure \ref{fig:huge} (a)-(c) provides an example of the weights distribution for non-watermarked model, regular watermarked model, and watermarked model trained with our watermark hiding technique. 
When embedding a watermark without applying hiding technique (Figure \ref{fig:huge} (b)), the weights distribution is significantly different from the non-watermarked weights distribution (Figure \ref{fig:huge} (a)). Yet, our watermark hiding technique makes the weights distribution very similar to the non-watermarked weights distribution, as shown in Figure \ref{fig:huge} (c).

Figure \ref{fig:huge} (d) shows the watermark detection accuracy of property inference attack on \uchida, \deepsign~ and \ours~for Benchmark 1-\bits. The attacks perform very differently when the watermarked models are trained with or without our watermark hiding technique.
When the hiding technique is not applied, the property inference attack is extremely effective as the accuracy climbs above 95\% after 10 epochs of training for all three watermarking algorithms.
However, when the watermark hiding algorithm is applied, the performance of property inference attack drops dramatically. The detection accuracies rarely go above 60\%, which are only slightly better than random guess. 
Note that property inference attack has the worst performance on \deepsign~ when embedding watermarks without watermark hiding. We also observe that \deepsign~has the smallest impact on weights distribution compared with \uchida~and \ours. However, the watermark detection accuracy is still above 90\% for \deepsign, which implies that there are some invisible patterns exist in the watermarked weights distribution. Given enough watermarked samples, these patterns can be a property inference attack. On the other hand, our covertness regularizer $\mathcal{E}_{det}$ is designed specifically to hide the existence of watermarks, and is much more effective in defending against watermark detection. 
We refer to Appendix \ref{apdx:covertness} for similar attack results on other benchmarks. 

\begin{figure}
	\centering 
	\includegraphics[width=0.9\columnwidth]{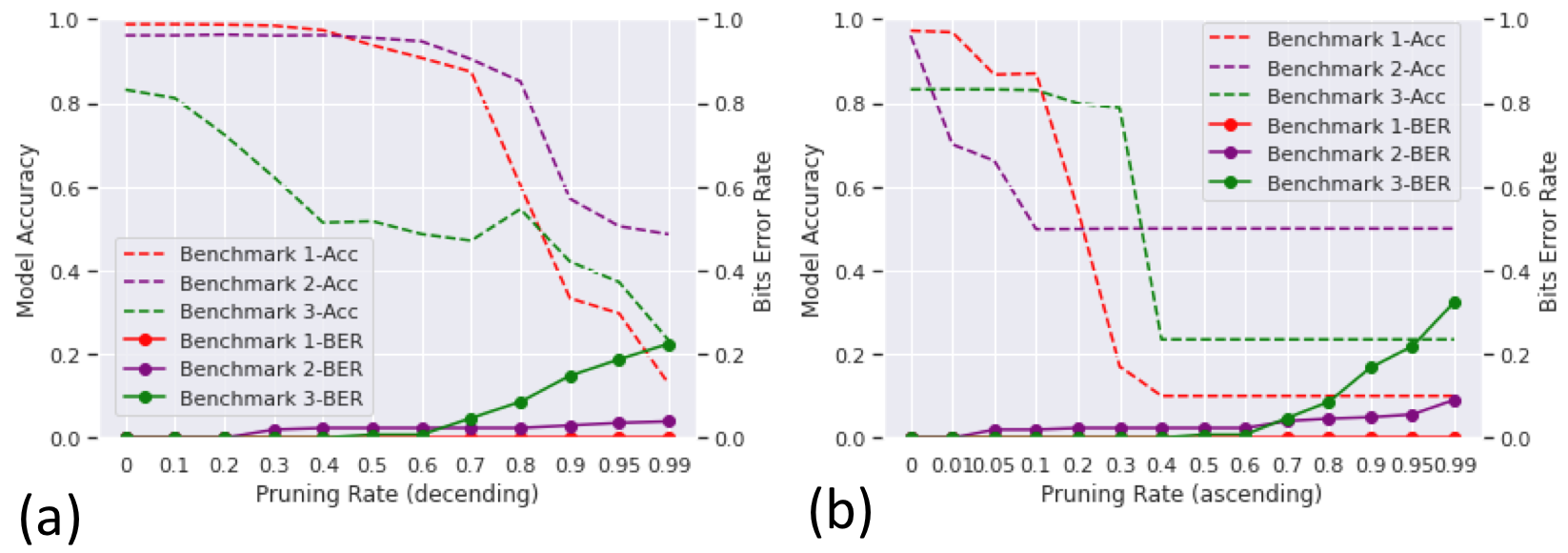}
	\caption{(a) Pruning attacks in descending order, (b) Pruning attacks in ascending order.}
	\label{fig:pruning-comb} 
\end{figure}



\subsection{Robustness}
\label{sec:eval-robustness}

\subsubsection{Robustness against Overwriting Attacks}
\label{sec:eval-robustness-overwriting}
We attempt to overwrite the original watermark with new watermarks of either the same size as the original watermark (i.e. a 256-bit binary string or a $128 \times 128$ 3-channel logo image), or of a larger size (i.e. a 512-bit binary string or a $256 \times 256$ 3-channel logo image). 
Figure \ref{fig:huge} (e) shows the embedding loss of overwriting attack on \bits watermark for all three benchmarks. In all cases, we can see that there are only tiny increases in the embedding losses. The BER is 0 for all cases in Figure \ref{fig:huge} (e). The results show that overwriting by a larger watermark does not cause larger embedding losses after convergence, although the embedding losses increase faster at the beginning of the attack. The attack results on \img~watermark are similar and we leave it to Appendix \ref{apdx:robustness}. We also refer to Appendix \ref{apdx:robustness} for the results of comparison amongst \ours, \uchida~ and \deepsign~against overwriting attacks. 

An attacker may also try to remove the watermark by embedding new watermarks into the neural networks with a different watermarking algorithm. 
However, we find that this approach is not effective for any combinations of the existing watermarking algorithms, and the resulting plots are very similar to Figure \ref{fig:huge} (e). 

\begin{table*}[]
\begin{tabular}{|c|c|c|c|c|c|c|c|c|}
\hline
\textbf{BM}        & \textbf{Type}      & \textbf{WM} & \textbf{Adi et al.} & \textbf{REFIT (LR=0.05)} & \textbf{REFIT (LR=0.04)} & \textbf{REFIT (LR=0.03)} & \textbf{REFIT (LR=0.02)} & \textbf{REFIT (LR=0.01)} \\ \hline
\multirow{4}{*}{1} & \multirow{2}{*}{FTAL} & BITS             & N/A                 & 10.10\%                  & 10.02\%                  & 10.12\%                  & N/A                      & N/A                      \\ \cline{3-9} 
                   &                       & IMG              & N/A                 & 8.92\%                   & 9.42\%                   & 8.92\%                   & N/A                      & N/A                      \\ \cline{2-9} 
                   & \multirow{2}{*}{RTAL} & BITS             & N/A                 & 10.10\%                  & 10.12\%                  & 9.52\%                   & N/A                      & N/A                      \\ \cline{3-9} 
                   &                       & IMG              & N/A                 & 8.92\%                   & 8.92\%                   & 8.92\%                   & N/A                      & N/A                      \\ \hline
\end{tabular}
\caption{Results of fine-tuning attack on Benchmark 1 with learning rate schedule in \citet{adi18} and REFIT. Each percentage represents the model accuracy when BER is greater than 15\% or embedding loss is greater than 0.1.}
\label{tb:finetune-maintext}
\end{table*}

\subsubsection{Robustness against Fine-tuning Attacks}
\label{sec:eval-robustness-ft}
REFIT \cite{chen2019refit} does not provide a concrete method to select the optimal learning rate that can remove the watermark while maintaining model performance. Nevertheless, we search such an optimal learning rate by varying the magnitude of the learning rate during fine-tuning and see its effect on watermark accuracy. Specifically, starting from the original learning rate (0.0001 for all three benchmarks), the learning rate is doubled every 10 epochs in the fine-tuning process.   
Figure \ref{fig:huge} (g) presents the model accuracy and watermark BER of this process for all three benchmarks with the \bits~ watermark. We observe that when the learning rate is large enough, e.g. after 40 epochs for Benchmark 1 and 3, or after 60 epochs for Benchmark 2, the model accuracy will significantly decrease at the beginning of each epoch when the learning rate is doubled. 
When the learning rate is moderately large, the model accuracy may gradually restore within the next 10 epochs. However, for these learning rates, there is no noticeable change in BER. 
When the learning rate is too large (e.g. after 60 epochs for Benchmark 1), the model will be completely destroyed and there will be no incentive for the attacker to redistribute it. 

The results in Figure \ref{fig:huge} (g) indicate that the optimal fine-tuning learning rates in REFIT that can remove \ours~ while also preserving model performance may not exist or cannot be found easily. To further demonstrate \ours's robustness against fine-tuning attacks, we evaluate REFIT with initial learning rates range from 0.01 to 0.05 as typically used in \citet{chen2019refit}. In Table \ref{tb:finetune-maintext}, we show the model accuracy when the watermarks are \emph{moderately removed} in fine-tuning attacks on Benchmark 1. That is, for \bits~ watermark we record the model accuracy when BER is greater than 15\% or the embedding loss is greater than 0.1. For \img~ watermark we record the model accuracy when the embedding loss is greater than 0.1. We denote ``N/A'' if the BER or embedding loss is never above the threshold during the entire fine-tuning process. As shown in Table \ref{tb:finetune-maintext}, for all cases, the model performance is significantly degraded when the watermarks are moderately removed. The attack results on Benchmark 2 and 3 are very similar and we refer to Appendix \ref{apdx:robustness} for the results. 

We have also tried fine-tuning the target model with a different dataset (transfer learning case in \citet{chen2019refit}), however we observe that it does not provide additional gain in watermark removal, but causes much larger decrease in model performance.

\subsubsection{Robustness against Weights Pruning Attack}
\label{sec:eval-robustness-pruning}
We use the weights pruning approach \cite{SongHan} to compress our watermarked deep learning model $\target$. Specifically, we set $\alpha$\% of the parameters in $w$ with the smallest absolute values to zeros. Figure \ref{fig:pruning-comb} (a) shows pruning's impact on watermark accuracies for all three benchmarks on \bits. For Benchmark 1-\bits~and 2-\bits, \ours~can tolerate up to 99\% pruning ratio. For Benchmark 3-\bits, the BER will be around 20\% when the pruning ratio is 95\% but still far less than 50\%, while the watermarked model accuracy has already dropped to around 40\%. Given \ours's strong robustness when we prune parameters from small to large, we also test the case when we prune parameters from large to small (Figure \ref{fig:pruning-comb} (b)). Surprisingly, while the model accuracy drops more quickly, 
the watermarks remain robust against pruning. The BER is still negligible for Benchmark 1 and 2 even when we pruned for over 95\% parameters. This implies that \ours~does not embed too much watermark information on any particular subset of the weights. Figure \ref{fig:huge} (f) shows the extracted image watermark for Benchmark 1-\img~ after weights pruning attack with different pruning ratios. 
Even with a pruning ratio of $95\%$, the logo image can still be clearly recognized, while the pruned model suffers a great accuracy loss.

\section{Conclusions}
\label{sec:conclusions}
In this work we generalize existing white-box watermarking algorithms for DNN models. We first outlined a theoretical connection to the previous work on white-box watermarking for DNN models. We present a new white-box watermarking algorithm, \ours, whose watermark extracting function is also a DNN and which is trained using an adversarial network. We performed all plausible watermark detection and removal attacks on \ours, and showed that \ours~ is both more covert and robust compared to the existing works. For future work, we would like to develop a theoretical boundary for how much similarity between embedded and extracted watermark message is needed for a party can claim ownership of the model.

\bibliographystyle{ACM-Reference-Format}
\bibliography{ref}

\appendix

\section{Additional Requirements for White-box Watermarks}
\label{apdx:requirements}

In the literature \cite{adi18,zheng19} further properties of watermarking algorithms have been defined. We review them and show that they are met by our new watermarking scheme here. 

\textbf{Non-trivial ownership}: This property requires that an adversary is not capable of producing a key that will result on a predictable message for any DNN. 
Formally, $\forall \key \in \mathbb{K}$, we have
$$
\Pr_{w \in \pspace, m \in \mathbb{M}}[\textbf{Extract}(w, \key) = m] \approx \frac{1}{|\mathbb{M}|}
$$
If this requirement is not enforced, an attacker can find a $\key$ that can extract watermark message $m$ from any $w \in \pspace$, and then he/she can claim ownership of any DNN models.
We require any {\em valid} extractor to prevent this attack. We show the validity of \ours~ in Appendix \ref{sec:integrity}. 

\textbf{Unforgeability}: This property requires that an adversary is not capable of reproducing the key for a given watermarked model. 
Formally, $\forall w, w' \in \pspace$ and $\forall m \in \mathbb{M}$, we have 
$$
\Pr_{\key \in \mathbb{K}}[\textbf{Extract}(w, \key)=m] = 
\Pr_{\key \in \mathbb{K}}[\textbf{Extract}(w', \key)=m]
$$
Intuitively, this requirement implies that an adversarial cannot learn anything about $\key$ from model weights $w$. 
This property can be easily achieved by the owner cryptographically committing to and timestamping the key \cite{adi18} and is orthogonal to the watermarking algorithms described in this paper.

\textbf{Ownership Piracy}: 
This property requires that an adversary that embeds a new watermark into a DNN does not remove any existing ones. We show that this property holds in Section \ref{sec:eval-robustness-overwriting} where we evaluate \ours~'s robustness against overwriting attack.

\section{DeepSigns Watermarking Scheme}
\label{sec:rouhani}
In the DeepSigns scheme \cite{Rouhani}, Rouhani et al.~replace the feature selection part in their watermarking algorithm compared to Uchida et al.'s scheme.
The features of $w$ they choose to embed the watermark into are the activations of a chosen layer of the DNN given a trigger set input. Hence the feature extraction key space $\mathbb{K}_{FE}$ is a product space of a matrix space and input space $X$. 
The feature extraction key is $\fekey=(\fematrix, x)$. 
\footnote{DeepSigns assumes a Gaussian Mixture Model (GMM) as the prior probability distribution (pdf) for the activation maps. In the paper, two WM-specific regularization terms are incorporated during DNN training to align the activations and encode the WM information. We only describe the principle of the watermarking algorithm in this paper, but implement their precise algorithm for our generic detection attack.}
\begin{itemize}
	\item $\ftwm:\pspace \times \mathbb{K}_{FE} \rightarrow \pspace_{l}$ where 
	$\ftwm(w, (\fematrix, x))$ outputs the activations of the selected layer of the DNN given trigger set $x \subseteq X$.
	\item $\fext$, $\dist$ are the same as in Uchida et al.'s scheme.
\end{itemize}

\section{RIGA's Watermarking Flowcharts}
\label{apdx:flowchart}

\begin{figure}[!htb]
	\includegraphics[width=.45\textwidth]{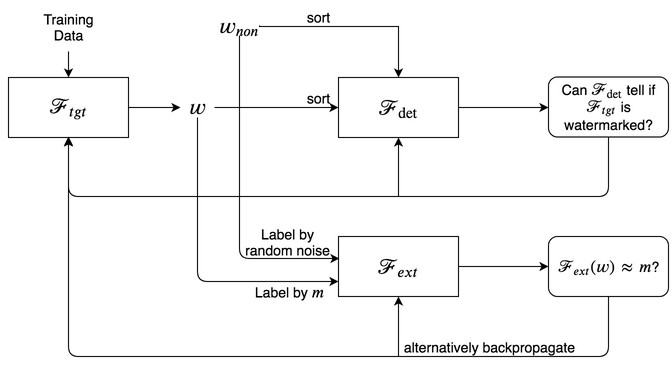}
	\label{fig:completeflow}
	\caption{Complete White-Box Watermarking Flowcharts for \ours.}
\end{figure} 

\section{Details of Evaluation Setup}

\subsection{Datasets} 
\label{apdx:dataset}
We evaluate \ours~ using three datasets: (1) the MNIST handwritten digit data, (2) the CelebFaces Attributes Dataset (CelebA) \cite{liu2018large} containing 202,599 face images of 10,177 identities with coarse alignment. We train deep learning models for gender recognition on this dataset, and (3) Amazon Fine Food Dataset \cite{amazonfinefood} containing 568,454 reviews of fine foods from amazon, and the corresponding ratings range from 1 to 5. We train natural language processing models for sentiment classification on this dataset.

\subsection{Models}
\label{apdx:model}
We implement several different target networks with varied complexities. Some of the networks are adapted from existing ones by adjusting the number of outputs of their last fully connected layer to our tasks. 
For the digit classification on MNIST (Benchmark 1), our target network consists of 2 convolutional layers and 2 batch normalization layers. For the gender recognition on CelebA (Benchmark 2), we use Inception-v3 adapted from \citet{szegedy2016rethinking}. For sentiment classification task on Amazon Fine Food (Benchmark 3), our target network consists of a Long Short-Term Memory (LSTM) layer and a fully-connected output layer. 

\subsection{Property Inference Attack}
\label{apdx:pi_attack-setup}
For each of \uchida, \deepsign~and \ours~watermarking algorithms, with or without using our watermark hiding technique, we train 1024 watermarked models and 1024 non-watermarked models to perform property inference attack. All of the models have the exact same architectures and all trained using the same datasets as the benchmarks. We also generate 100 non-watermarked and 100 watermarked models by each of the watermarking algorithms as test sets. We construct the training set for $\detector$ by extracting the watermarking features for each model and label it according to whether or not it is watermarked. 
All of the generated models are well-trained and watermarks are well embedded. 

\section{Additional Evaluation Results}

\subsection{Covertness}
\label{apdx:covertness}
\begin{figure}
	\centering 
	\subfigure[]{
	\includegraphics[width=0.45\columnwidth]{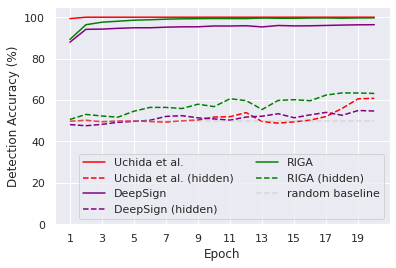}
	\label{fig:pi_attack-2}}
	\subfigure[]{
	\includegraphics[width=0.45\columnwidth]{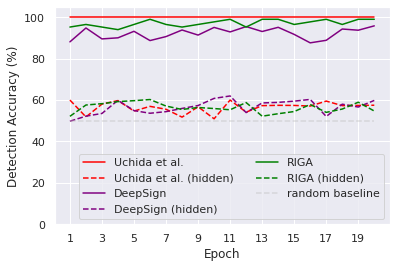}
	\label{fig:pi_attack-3}
	}
	\caption{Results of property inference attacks on (a) Benchmark 2, and (b) Benchmark 3. }
\end{figure}

Figure \ref{fig:pi_attack-2} and \ref{fig:pi_attack-3} show the results for property inference attack on Benchmark 2 and 3. Similar to the attack results on Benchmark 1, our watermark hiding technique significantly improves the covertness of watermarks.

\subsection{Robustness}
\label{apdx:robustness}
\begin{figure}
	\subfigure[]{
	\includegraphics[width=0.45\columnwidth]{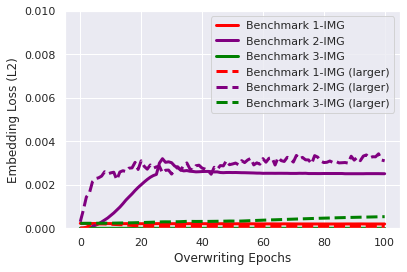}
	\label{fig:overwriting-img}
	}
	\subfigure[]{
	\includegraphics[width=0.45\columnwidth]{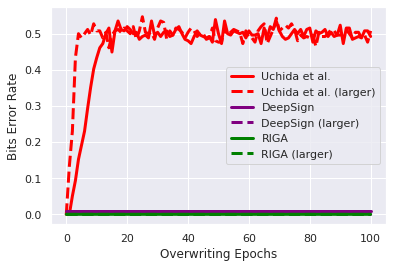}
	\label{fig:overwriting-comparison}
	}
	\caption{Results of (a) overwriting attack on \img, and (b) comparison of overwriting attacks on \uchida, \deepsign~and \ours~on Benchmark 1-\bits. }
\end{figure}

Figure \ref{fig:overwriting-img} shows the embedding loss of overwriting attack on \img~ watermark for all three benchmarks. Similar to the results on \bits~watermark, there are only tiny increases in the embedding losses for all cases. The results show that overwriting by a larger watermark does not cause larger embedding losses after convergence, although the embedding losses increase faster at the beginning of the attack.

Figure \ref{fig:overwriting-comparison} shows the comparison of \ours~ with \uchida~ and \deepsign~against overwriting attacks with Benchmark 1-\bits. We can see that both \deepsign~ and \ours~are robust against overwriting attacks, and the watermarks embedded by \uchida~will be removed as shown in the literature \cite{TWang}. We conjecture that this is because for both \deepsign~and \ours, the secret keys $\fekey$ and $\mekey$ are complex enough such that there is a negligible probability that a randomly generated key will be similar to the original key.

\begin{figure}
	\centering
	\includegraphics[width=0.75\columnwidth]{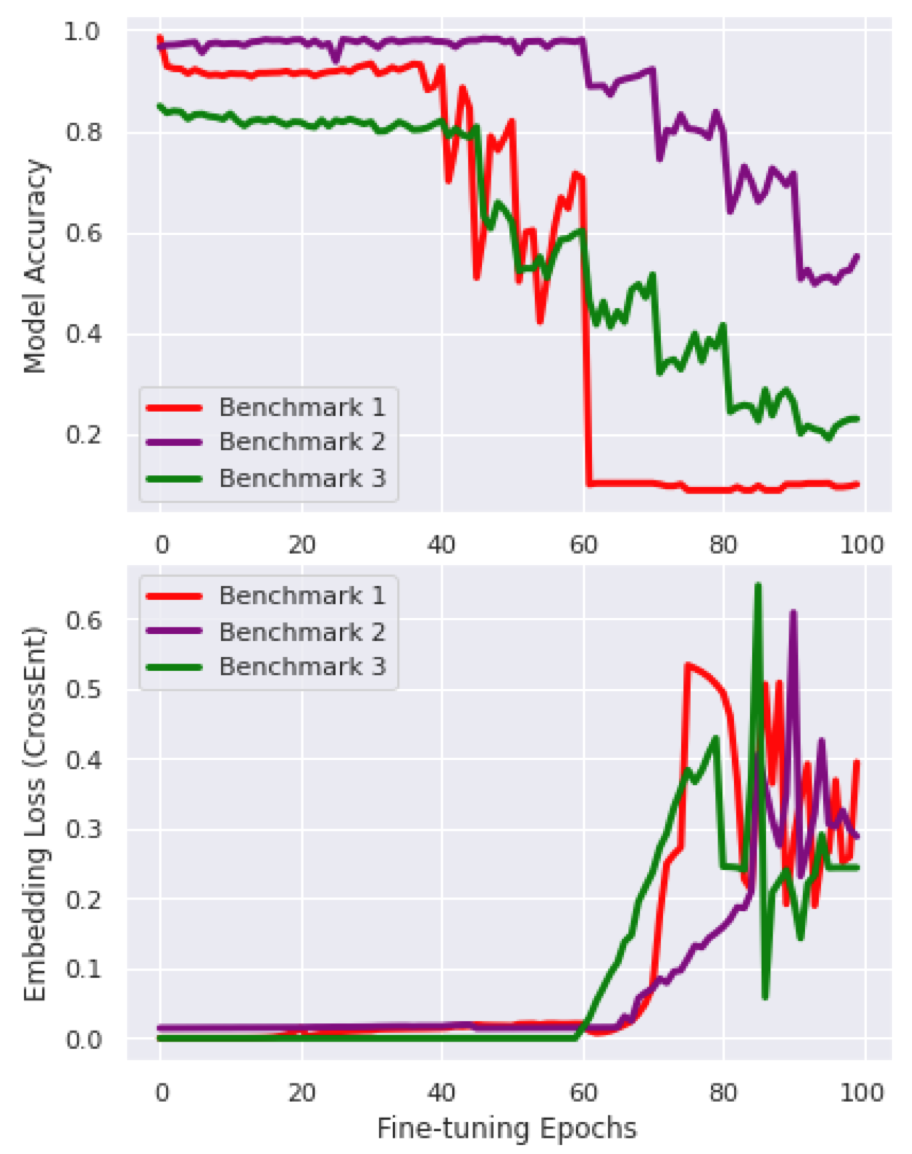}
	\caption{Training curves for finding the optimal learning rates for watermark removal with the metric of embedding loss. }
	\label{fig:finetune-loss}
\end{figure}

Figure \ref{fig:finetune-loss} shows the variation of watermark embedding loss during the optimal learning rate searching process in Section \ref{sec:eval-robustness-ft} for \bits~ watermark. Similar to the results for BER metric, we observe that when the learning rate is moderately large, the model accuracy may gradually restore within the next 10 epochs. However, for these learning rates, there are only small increase in the embedding loss. When the learning rate is too large, the model will be completely destroyed and there will be no incentive for the attacker to redistribute it. 

Table \ref{tb:finetune-maintext} shows the results of fine-tuning attacks on Benchmark 2 and 3, where we record the model accuracy when watermarks are moderately removed. Similar to the results on Benchmark 1, for all cases, the model performance decreases significantly when the watermarks are moderately removed, which indicate that it is very difficult to remove \ours~while maintaining model performance by using a fine-tuning based removal attack.

\subsection{Fine-pruning}
\label{sec:fine-pruning}
Fine-pruning is another effective removal algorithm for black-box watermarks \cite{liu2018fine}. However, \citet{chen2019refit} report that pruning is not necessary with a properly designed learning rate schedule for fine-tuning even for black-box watermarks. For the completeness of evaluation, we also evaluate the performance of fine-pruning on \ours. Fine-pruning first prunes part of the neurons that are activated the least for the training data, and then performs fine-tuning. Following the settings in \citet{liu2018fine}, we keep increasing the pruning rate until the decrease of model accuracy becomes noticeable, and then apply fine-tuning where the learning rate schedule either follows Adi et al. or REFIT. Table \ref{tb:finepruning} presents the results for \bits~watermarks and FTAL. For all three benchmarks, we find that the results are roughly similar to fine-tuning without pruning. Therefore, we conclude that fine-pruning is also not an effective removal attack against \ours. 

\begin{table*}[]
\begin{tabular}{|c|c|c|c|c|c|c|}
\hline
\textbf{Benchmark} & \textbf{Adi et al.} & \textbf{REFIT (LR=0.05)} & \textbf{REFIT (LR=0.04)} & \textbf{REFIT (LR=0.03)} & \textbf{REFIT (LR=0.02)} & \textbf{REFIT (LR=0.01)} \\ \hline
1 - BITS           & N/A                    & 10.22\%                  & 10.84\%                  & 8.92\%                   & N/A                      & N/A                      \\ \hline
2 - BITS           & N/A                    & 50.00\%                  & 51.64\%                  & 50.00\%                  & 50.96\%                  & N/A                      \\ \hline
3 - BITS           & N/A                    & 22.86\%                  & 22.36\%                  & N/A                      & N/A                      & N/A                      \\ \hline
\end{tabular}
\caption{Results of fine-pruning attacks on \ours with \bits~watermark and FTAL.}
\label{tb:finepruning}
\end{table*}

\subsection{Validity}
\label{sec:integrity}
Validity, or non-trivial ownership, requires that the ownership of a non-watermarked model is not falsely assumed by the watermark extraction algorithm. 
If an owner tries to extract a watermark from a non-watermarked model, the extracted message must be different with overwhelming probability to satisfy the validity requirement. We evaluate the worst scenario to demonstrate the validity of our proposed scheme:

\begin{itemize}
	\item Alice embeds watermark $m$ into target model $\mathcal{F}_{tgt_1}$ with $\mathcal{F}_{ext_1}$. 
	\item Bob embeds same watermark $m$ into target model $\mathcal{F}_{tgt_2}$ with  $\mathcal{F}_{ext_2}$. 
	\item $\mathcal{F}_{tgt_1}$ and $\mathcal{F}_{tgt_2}$ have the exact same architectures, trained with exact same dataset, hyperparameters and optimizing algorithm. 
	\item $\mathcal{F}_{ext_1}$ and $\mathcal{F}_{ext_2}$ have the exact same architectures, trained with the exact same hyperparameters and optimizing methods. 
\end{itemize}

\begin{figure}
	\centering
	\subfigure[Original]{
		\includegraphics[width=0.25\columnwidth]{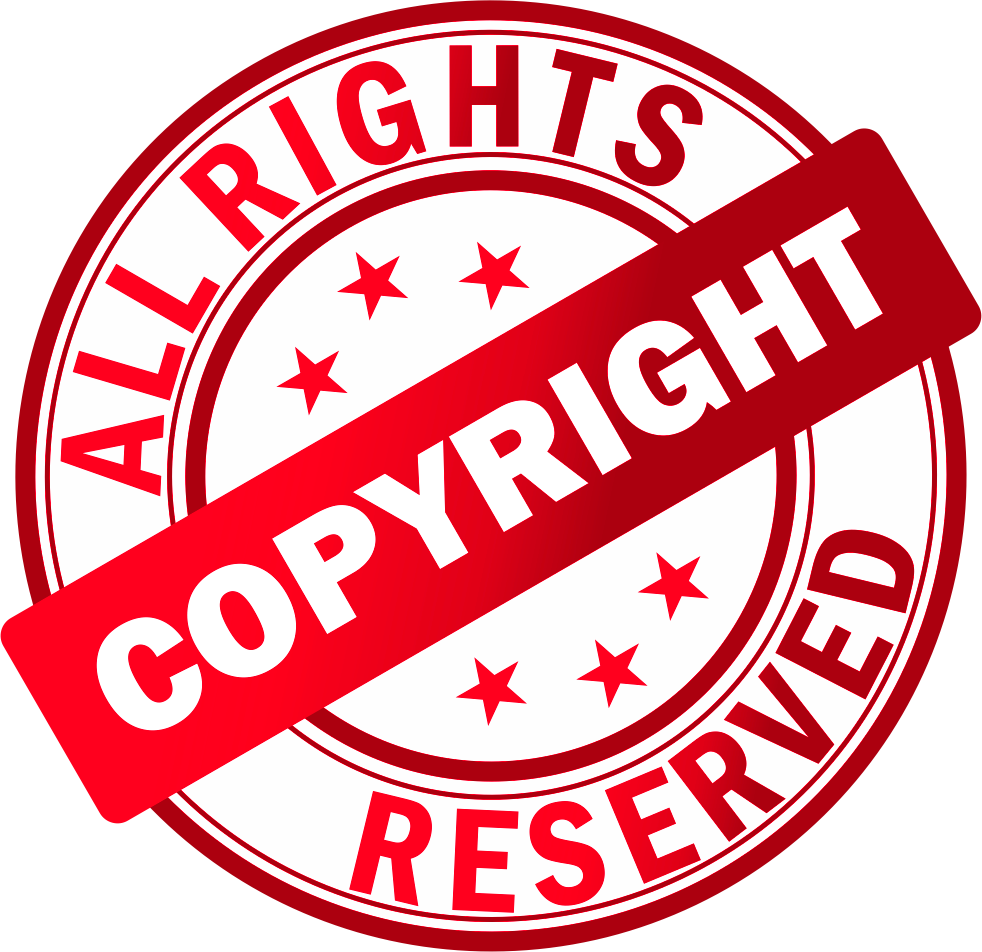}}
	\subfigure[$\mathcal{F}_{ext_1}(\mathcal{F}_{tgt_1})$]{
		\includegraphics[width=0.25\columnwidth]{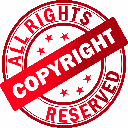}}
	\subfigure[$\mathcal{F}_{ext_1}(\mathcal{F}_{tgt_2})$]{
		\includegraphics[width=0.25\columnwidth]{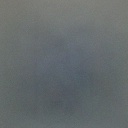}}
	\caption{Results of verifying the validity of RIGA under worst case. (a) is the original watermark image, (b) is the watermark image extracted by Alice from her own model, (c) is the watermark image extracted by Alice from Bob's model. }
	\label{fig:integrity} 
\end{figure}

We perform the above setup and test whether or not Alice can extract $m$ from $\mathcal{F}_{tgt_2}$ by using $\mathcal{F}_{ext_1}$. 
Figure \ref{fig:integrity} shows the results when the watermark $m$ is a logo image. 
As shown in Figure \ref{fig:integrity} (c), $\mathcal{F}_{ext_1}$ can only extract an extremely blurred image where the logo is extremely difficult, if at all, to recognize. 

\begin{table*}[]
\begin{tabular}{|c|c|c|c|c|c|c|c|c|}
\hline
\textbf{BM}        & \textbf{Type}      & \textbf{WM} & \textbf{Adi et al.} & \textbf{REFIT (LR=0.05)} & \textbf{REFIT (LR=0.04)} & \textbf{REFIT (LR=0.03)} & \textbf{REFIT (LR=0.02)} & \textbf{REFIT (LR=0.01)} \\ \hline
\multirow{4}{*}{2} & \multirow{2}{*}{FTAL} & BITS             & N/A                 & 50.00\%                  & 50.80\%                  & 51.80\%                  & 49.84\%                  & N/A                      \\ \cline{3-9} 
                   &                       & IMG              & N/A                 & 50.00\%                  & 51.40\%                  & 50.24\%                  & 49.54\%                  & N/A                      \\ \cline{2-9} 
                   & \multirow{2}{*}{RTAL} & BITS             & N/A                 & 50.00\%                  & 49.88\%                  & 49.88\%                  & 51.88\%                  & N/A                      \\ \cline{3-9} 
                   &                       & IMG              & N/A                 & 50.00\%                  & 50.00\%                  & 50.00\%                  & 50.36\%                  & N/A                      \\ \hline
\multirow{4}{*}{3} & \multirow{2}{*}{FTAL} & BITS             & N/A                 & 23.48\%                  & 21.12\%                  & N/A                      & N/A                      & N/A                      \\ \cline{3-9} 
                   &                       & IMG              & N/A                 & 21.82\%                  & 22.09\%                  & N/A                      & N/A                      & N/A                      \\ \cline{2-9} 
                   & \multirow{2}{*}{RTAL} & BITS             & N/A                 & 21.22\%                  & 19.80\%                  & N/A                      & N/A                      & N/A                      \\ \cline{3-9} 
                   &                       & IMG              & N/A                 & 20.92\%                  & 19.80\%                  & N/A                      & N/A                      & N/A                      \\ \hline
\end{tabular}
\caption{Results of fine-tuning with learning rate schedule in \citet{adi18} and REFIT. Each percentage represents the model accuracy when BER is greater than 15\% or embedding loss is greater than 0.1.}
\label{tb:finetune-appendix}
\end{table*}

\end{document}